\documentclass[aps,prl,twocolumn,showpacs,preprintnumbers,amsmath,amssymb,superscriptaddress]{revtex4-2}

\usepackage{color,soul}

\usepackage{graphicx}
\usepackage{dcolumn}
\usepackage{bm}
\newcommand{\nc}{\newcommand}
\nc{\be}{\begin{equation}}
\nc{\ee}{\end{equation}}
\nc{\bal}{\begin{align}}
\nc{\eal}{\end{align}}
\nc{\bea}{\begin{eqnarray}}
\nc{\eea}{\end{eqnarray}}
\nc{\bean}{\begin{eqnarray*}}
\nc{\eean}{\end{eqnarray*}}
\nc{\mb}{\mbox}
\nc{\rnc}{\renewcommand}
\nc{\vk}{\mb{\bf k}}
\nc{\vp}{\mb{\bf p}}
\nc{\vn}{\mb{\bf n}}
\nc{\vq}{\mb{\bf q}}
\nc{\rr}{\mb{\bf r}}
\nc{\vz}{\hat {\mb{\bf z}}}
\nc{\vj}{\mb{\boldmath$j$}}
\nc{\vg}{\mb{\boldmath$g$}}
\nc{\x}{\mb{\boldmath$x$}}
\nc{\A}{\mb{\boldmath$A$}}
\nc{\va}{\mb{\boldmath$a$}}
\nc{\vs}{\mb{\boldmath$\sigma$}}
\nc{\vpi}{\mb{\boldmath$\pi$}}
\nc{\nab}{\nabla}
\nc{\X}{\sf x}
\nc{\kk}{{\bf k}}
\nc{\pp}{{\bf p}}
\nc{\qq}{{\bf q}}
\nc{\Qq}{{\bf Q}}
\nc{\vl}{{\bf l}}
\nc{\Kk}{{\bf K}}
\nc{\upspin}{{\uparrow}}
\nc{\dspin}{{\downarrow}}
\nc{\vecq}{{\bf q}}
\nc{\veck}{{\bf k}}
\nc{\vecp}{{\bf p}}
\nc{\vecl}{{\bf l}}
\nc{\vecr}{{\bf r}}
\nc{\vecx}{{\bf x}}
\nc{\vecR}{{\bf R}}
\nc{\vecG}{{\bf G}}
\nc{\vecA}{{\bf A}}
\nc{\vecpi}{{\bf \pi}}
\nc{\vecL}{{\bf L}}
\nc{\vecK}{{\bf K}}
\newcommand{\ket}[1]{| #1 \rangle}
\newcommand{\bra}[1]{\langle #1 |}

\usepackage{graphicx}
\usepackage{txfonts}
\usepackage{dcolumn}
\usepackage{bm}
\usepackage{hyperref}
\hypersetup{colorlinks,citecolor=red,filecolor=black,linkcolor=blue,urlcolor=blue}
\usepackage{lipsum}
\usepackage{subcaption}
\usepackage{subfloat}
\usepackage{caption}
\usepackage{natbib}

\usepackage[utf8]{inputenc}
\usepackage{calc}
\usepackage{accents}

\nc{\argg}{\text{Arg}}
\nc{\bd}{\textbf}
\nc{\bds}{\boldsymbol}
\nc{\ham}{\hat{\mathcal{H}}}
\nc{\im}{\text{Im}}
\nc{\la}{\langle}
\nc{\ra}{\rangle}
\nc{\re}{\text{Re}}
\nc{\rn}[1]{%
	\textup{\uppercase\expandafter{\romannumeral#1}}%
}
\nc{\sgn}{\text{Sgn}}
\nc{\tit}{\textit}
\nc{\tr}{\text{Tr}}

\nc{\les}{\leqslant}
\nc{\ges}{\geqslant}

\def\ket#1{{\left|#1\right\rangle}}
\def\bra#1{{\left\langle #1 \right|}}

\def\be{\begin{eqnarray}}
\def\ee{\end{eqnarray}}

\usepackage{stackengine}

\begin{document}
\setstcolor{red}

\title{Effects of Coupling Between Chiral Vibrations and Spins in Molecular Magnets}
\author{Aman Ullah}
\affiliation{Department of Physics, University of Nevada, Reno, NV 89557, USA}
\author{Sergey A. Varganov}
\affiliation{Department of Chemistry, University of Nevada, Reno, NV 89557, USA}
\author{Yafis Barlas}
\affiliation{Department of Physics, University of Nevada, Reno, NV 89557, USA}
\email{ybarlas@unr.edu}

\begin{abstract}
In single molecular magnets, chiral vibrations carrying vibrational angular momentum ($\hat{L}^{\text{vib}}$) emerge due to the splitting of a doubly degenerate vibrational mode. Here, we identify a new type of effective spin-vibrational coupling responsible for lifting this degeneracy, which can facilitate optically selective excitations. In the presence of an external Zeeman field, this coupling breaks both inversion (in-plane parity) $\mathcal{P}$ and time-reversal $\mathcal{T}$ symmetries, imparting distinct geometric phases to the resulting dressed spin-vibronic states. The wave function of the spin-vibronic state is characterized by a $\pi$-Berry phase, which results in magneto-optical circular dichroism. 
This framework is validated using density functional theory and multi-reference \emph{ab initio} calculations on the Ce(trenovan) molecular magnet.
\end{abstract}

\maketitle

Chirality, arising from the interplay of spin, lattice, and charge degrees of freedom in quantum materials, is emerging as a key feature that enables the control and manipulation of quantum states~\cite{ueda2023chiral, wang2024chiral, aiello2022chirality}. Multiple degenerate vibrational modes ($\omega_{\pm}$) can exist in systems with point group symmetries that allow two-dimensional $E$-irreducible representations. Due to symmetry-breaking perturbations, these modes acquire distinct vibrational-angular momentum ($\hat{L}^{vib}$) \cite{shushkov2024novel,coh2023classification, lujan2024spin,chaudhary2024giant}, corresponding to circularly polarized states of opposite handedness~\cite{chen2018chiral,zhang2015chiral}. However, the mechanisms underlying the lifting of their degeneracy and properties of the resulting dressed chiral states remain unresolved. In this Letter, we report on a new type of spin-vibrational coupling (SVC) that breaks both in-plane parity ($\mathcal{P}$) and time-reversal ($\mathcal{T}$) symmetry and lifts the degeneracy of the vibrational modes, and study their optical properties.

\begin{figure}[t]
\includegraphics[width=8.5cm,keepaspectratio]{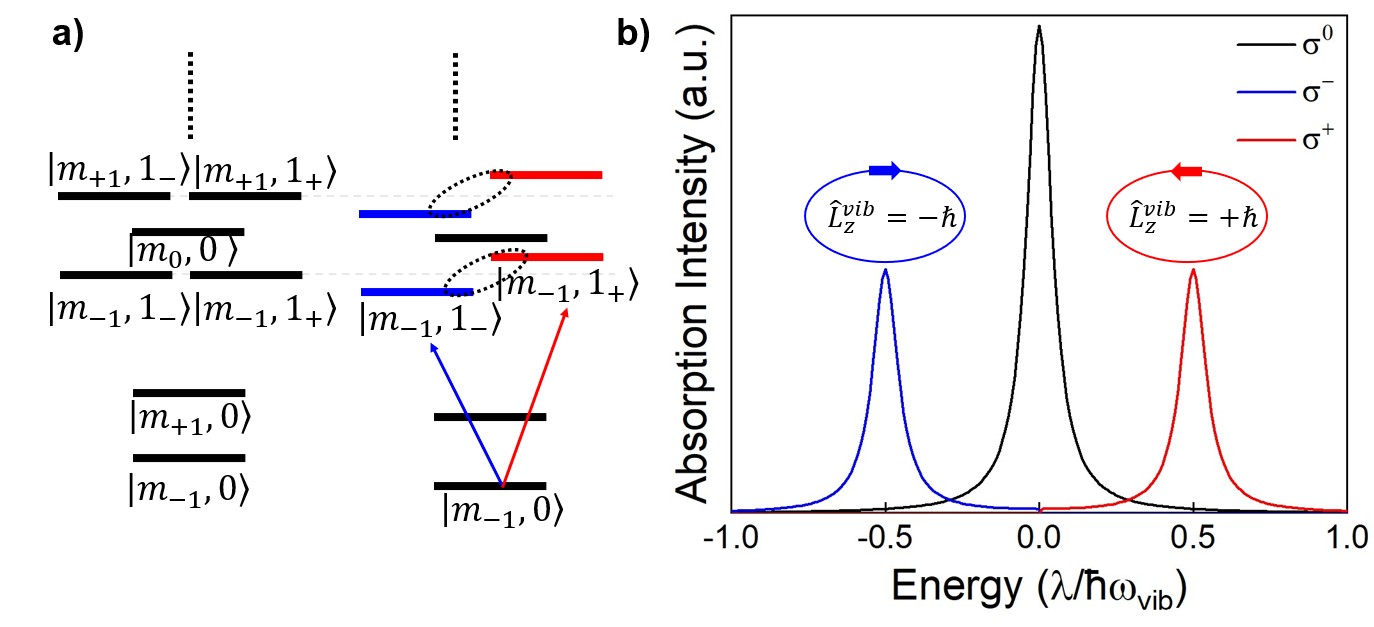}\\
\caption{\small Emergence of spin-vibronic states and their selective excitation.
\textbf{a)} In the absence of SVC, the system exhibits vibronic states (black horizontal lines). These couplings lift this degeneracy (blue and red horizontal lines), enabling angular momentum exchange between the split states (elliptical dashed curves). The vertical dashed line indicates higher-energy vibronic states. \textbf{b)} Without SVC, a single peak will appear at $\hbar \omega_{\text{vib}}$, preserving spherical symmetry. When SVC is introduced, this symmetry breaks, imparting angular momentum ($\pm \hbar$) that couples selectively to circularly polarized light ($\sigma^\pm$). The peaks associated with $\sigma^{\pm}$ split into two (red and blue), appearing at $\pm 0.5$ $\lambda$, where $\lambda$ is the SVC constant and $\hbar \omega_{\text{vib}}$ is the vibrational mode frequency.  We set $\hbar \omega_{\text{vib}}$ set to 0.}

\label{fig:1}
\end{figure}

Previous studies of chiral phonons have focused on linear SVC of the form $\hat{L}^{vib} \cdot \hat{J}$ \cite{zhang2014angular, hernandez2023observation}, which remains invariant under inversion symmetry. In single molecular magnets, due to crystal field effects, the SVC has the form $\hat{L}^{vib} O_k^q(\hat{J})$, where $O_k^q(\hat{J})$ are Stevens operators. Because the Stevens operators are polynomials in spin angular momentum, which incorporates both spin and orbital contributions, the even rank $O_k^q(\hat{J})$ are not invariant under inversion. 
Therefore, additional SVCs can be generated when inversion symmetry is broken in single molecular magnets, dressing the spins with vibrations and encoding them with new quantum numbers (see Fig.~\ref{fig:1} a).

The SVC Hamiltonian $H_{s-vib}$ is attained by analyzing the vibrational dependence of the spin Hamiltonian in the circular vibrational coordinates $\delta q_\pm$. When inversion symmetry is broken, this yields an effective 
``spin-vibrational-orbit" coupling that mediates angular momentum transfer between the vibronic states and lifts their degeneracy. The schematic energy levels in Fig.~\ref{fig:1} a) show the effect of the SVC on degenerate states. When these couplings are included, the degenerate vibronic states form pairs of dressed spin-vibronic states. In addition to lifting the degeneracy, the dressed spin-vibronic states acquire distinct geometric phases as their quantum numbers are modified. This is quantified by computing the Berry phase in the SVC parameter space, where the vibronic states $\ket{m_J,  1_\mp}$ form a conical intersection with $\pm \pi$ Berry phases \cite{berry1984quantal,yarkony1996diabolical, canali2003chern,valahu2023direct}. We show that due to the presence of $\pi$-phase, the dressed spin-vibronic states can be selectively excited in optical dichroism experiments, as indicated in Fig.~\ref{fig:1}b, with selection rules imposed by the chirality of the dressed states.

We apply our theoretical framework to the molecular nanomagnet Ce(trenovan) \cite{lucaccini2017electronic} with $C_3$ rotational symmetry. The Ln(trensal) family has been extensively studied as a promising candidate for molecular spin qubits \cite{dergachev2025ab}. We determine the vibrational energies using density functional theory (DFT), while multi-reference $\emph{ab initio}$ calculations reveal the energies of the spin states and SVC. We also calculate the topological and optical properties of the spin-vibronic states in Ce(trenovan), demonstrating the practical relevance of our theoretical framework. Our findings establish an understanding of symmetry breaking, geometric properties, and optical selection rules in spin-vibronic states in molecular magnets. They are also relevant for novel quantum applications, such as quantum sensing, where the sensitivity of spin-vibronic states to external fields can enhance precision measurements, and quantum initialization, and enable robust state preparation \cite{qi2011topological}.


\emph{Model}.---To model the SVC, we started with an effective Hamiltonian, $\hat{H}_{eff} = \hat{H}_s + \hat{H}_Z + \hat{H}_{vib}$, where $\hat{H}_s$ represents the spin Hamiltonian, $\hat{H}_Z$ is the Zeeman interaction,  $\hat{H}_{vib}$ describes the vibrational modes. Explicitly, $ \hat{H}_Z = \mu_B g_J  {\bf B} \cdot \hat{\bf J}$,  where $\hat{\bf J}=\hat{\bf L}+\hat{\bf S}$ denotes the total angular momentum (including orbital and spin contributions), $g_J$ is the gyromagnetic ratio, $\mu_B$ denotes the Bohr magneton, and ${\bf B}$ is the magnetic field. The vibrational Hamiltonian for two degenerate modes with energy $\hbar\omega$ is $\hat{H}_{vib}  = \hbar\omega (\hat{a}_+^\dagger \hat{a}_+ + \hat{a}_-^\dagger \hat{a}_-)$, where $\hat{a}_{\pm} =  (\hat{a} \pm i \hat{b})/\sqrt{2}$ are the bosonic operators for the circularly polarized basis for right- and left-handed vibrations. The bosonic operators $\hat{a} (\hat{b})$ correspond to the orthogonal mass and frequency normalized generalized displacements $\delta q_1 (\delta q_2)$ along the normal mode directions.

The SVC, which is the central focus of this paper, resulted from the dependence of $\hat{H}_s(q_+,q_-)$ on the generalized displacement of the degenerate vibrational modes $(\delta q_+,\delta q_-)$. 
The spin Hamiltonian can be expressed as: $\hat{H}_s= \sum_{k} \sum_{q=-k}^{k} B_k^q \hat{O}_k^q (J)$, where $B_k^q$ are crystal field parameters with rank $k$ and $q = -k, \cdots, k$ for the associated the Stevens  operator $\hat{O}_k^q (J)$~\cite{ungur2017ab}. The SVC Hamiltonian $\hat{H}_{s-vib}$ was attained by expanding $\hat{H}_s = \hat{H}^{0}_s + \hat{H}_{s-vib} $ about the equilibrium of the normal modes $q_i = q_{i0} + \delta q_i$ (see supplemental information (SI) for details). To second-order in $\delta q_i$, $\hat{H}_{s-vib}$ can be expressed as,
\begin{align}\label{eq:vb4}
\hat{H}_{s-\text{vib}} = -\frac{1}{4i}  \sum_{k=2,4,6} \sum_{q=-k}^{k} \lambda_{k,4}^{q} \hat{O}_k^q (J) \otimes (\hat{L}_+^{\text{vib}} - \hat{L}_-^{\text{vib}})+ \cdots,
\end{align}
where $\lambda_{k,4}^{q}$ are second-order SVC constants, and $\cdots$ denote vibrational mode coupling without angular momentum transfer. These additional terms, detailed in the SI, involve single- and double-excitations without angular momentum transfer and are therefore neglected in our discussion below. In Eq.~\eqref{eq:vb4}, the vibrational angular momentum operator, $\hat{L}_+^{\text{vib}}  = 2 \hat{a}_+^\dagger \hat{a}_-$ and $\hat{L}_-^{\text{vib}} = 2 \hat{a}_-^\dagger \hat{a}_+$ drive transitions in an effective two-level system with  $\hat{a}_+^\dagger |n_+, n_-\rangle = \sqrt{n_+ + 1} |n_+ + 1, n_-\rangle$ and $\hat{a}_- |n_+, n_-\rangle = \sqrt{n_-} |n_+, n_- - 1\rangle$.

Since, $\hat{\bf L}^{\text{vib}}$ is a pseudo-vector it transforms as $(\hat{L}_{x}^{\text{vib}},\hat{L}_{y}^{\text{vib}},\hat{L}_{z}^{\text{vib}}) \to (-\hat{L}_{x}^{\text{vib}},-\hat{L}_{y}^{\text{vib}},\hat{L}_{z}^{\text{vib}})$  under inversion. Therefore, in the presence of inversion symmetry, the coupling constants $\lambda_{k,4}^{q}$ in Eq.~\ref{eq:vb4} corresponding to the even parity Stevens operators ($q \in \textrm{even}$) must vanish. In contrast, the odd parity coupling constants $\lambda_{k,4}^{q}$ with $q \in \textrm{odd}$ can be non-zero. When inversion symmetry is broken, it generates a new type of SVC due to the presence of nonzero $\lambda_{k,4}^{q}$ with $q \in \textrm{even}$ coupling constants, leading to angular momentum transfer between the spin and vibrational degrees of freedom. Alternatively, while $\lambda_{k,4}^{q}$ with $q \in \textrm{odd}$ can be modified due to broken inversion symmetry, resulting in changes to the energy spectrum, they do not alter the quantum numbers of the eigenstates. To quadratic order in the vibrations ($\delta q_+, \delta q_-$), the coupling of $\hat{L}_{\pm}^{\text{vib}}$ with the spin $\hat{O}_k^q (J)$ via  $\lambda_{k}^q$, lifts the degeneracy of the right- and left-handed chiral vibration modes and induces an effective spin-vibrational-orbit interaction which results in angular momentum exchange, as we show below. 


\emph{Magneto-Optical Circular Dichorism}.--- To highlight the optical properties, we considered a spin system with $J=1$ and $m_J=-1,0,1$, setting diagonal element $B_2^0 = -1$ to favor $m_J = \pm 1$ ground states, and off-diagonal $B_2^{\pm 2} = 0.2$ to mix $m_J$ levels, and $\mu_B g_J \mathbf{B} / \hbar \omega_{\text{vib}} = 0.2$ to break time-reversal symmetry between $m_J = \pm 1$. The SVC, according to Eq.\eqref{eq:vb4}, contains contributions from first-order couplings $\partial B_k^q / \partial q_{a(b)}$, and second-order couplings $\lambda_{k,4}^q \propto \partial B_k^q / (\partial q_{a}\partial q_{b})$. For real systems, these parameters are determined using \emph{ab initio} methods. Normalized by $\hbar \omega_{\text{vib}}$, the vibronic spectrum (Fig.~\ref{fig:1}a) reveals topological features, with details of the calculations provided in the SI.

Selective excitation of spin-vibronic states is determined by SVC, their quantum numbers, and the geometric phase acquired in the SVC Hamiltonian's parameter space. The effective Hamiltonian for the ground state $\ket{0}$ and the two excited degenerate vibrational levels $\ket{1}$ and $\ket{2}$  is expressed as $\hat{H}_{\text{eff}} = \text{diag}\{0, \omega_{\text{vib}}, \omega_{\text{vib}}\} + \eta_1 \hat{\mathcal{L}}_+ + \eta_2 \hat{\mathcal{L}}_-$, where $\hat{\mathcal{L}}_+ = \ket{2}\bra{1}$ and $\hat{\mathcal{L}}_- = \ket{1}\bra{2}$ are the raising and lowering pseudo-angular momentum operators acting on the dressed spin-vibronic states. The coefficients $\eta_{1(2)}$ represent the effective SVC ($= \lambda_{k,4}^{q}$), modulated by the vibrational coordinates $(q_1,q_2)$ of the right (1)- and left (2)-handed modes. The corresponding eigenfunctions are $\mathbf{v}_0^T = (1, 0, 0)$ for the ground state and $\mathbf{v}_{\pm}^T = 1/\sqrt{2} (0, 1, \pm e^{i \phi})$ for the excited states, where $\phi = \tan^{-1}(\eta_2 / \eta_1)$. The eigenvalues are $E_0 = 0$ for the ground state and $E_{\pm} = \omega_{\text{vib}} \pm \sqrt{\eta_1^2 + \eta_2^2}$ for the excited states.

The system interacts with circularly polarized light, which drives transitions between the vibronic states. Here, $\sigma^+ = \ket{\mathbf{v}_+}\bra{\mathbf{v}_0}$ and $\sigma^- = \ket{\mathbf{v}_-}\bra{\mathbf{v}_0}$ are the circular polarization operators in the vibronic eigenbasis, representing dipole interactions that raise the angular momentum by $+1$ and $-1$, respectively, coupling the ground state $\mathbf{v}_0$ to the excited states $\mathbf{v}_+$ and $\mathbf{v}_-$. Their adjoints, $(\sigma^+)^\dagger = \ket{\mathbf{v}_0}\bra{\mathbf{v}_+}$ and $(\sigma^-)^\dagger = \ket{\mathbf{v}_0}\bra{\mathbf{v}_-}$, facilitate back transitions from the excited states to the ground state via stimulated emission. The intensity of these transitions depends on the eigenfunctions and the incident frequency ($\omega$), given by the intensity ($\text{I}$), $\text{I}_{\sigma^\pm} \propto |\langle \mathbf{v}_\pm | \sigma^\pm | \mathbf{v}_0 \rangle|^2$ for absorption, and $\text{I}_{(\sigma^\pm)^\dagger} \propto |\langle \mathbf{v}_0 | (\sigma^\pm)^\dagger | \mathbf{v}_\pm \rangle|^2$ for emission. This selective excitation, illustrated in Fig. \ref{fig:1}b, enables chiral light-matter interactions and provides a mechanism for quantum control. In magneto-optical experiments, such interactions can lead to observable effects like circular dichroism, where absorption differs for left- and right-handed circular polarization \cite{ji2025circularly, elmers2025chirality, tikhonov2022pump, huang2023circularly, baydin2022magnetic}.

\emph{Geometric properties}.---The spin-vibronic states exhibit distinct geometrical phases, which were determined by their Berry phases acquired in the two-dimensional parameter space $[\eta_+, \eta_-]$, where $\eta_+$ and $\eta_-$ are proportional to the coupling constants for right- and left-handed circularly polarized vibrational modes, respectively. The effective range of these displacements is set by the molecular potential energy surface, typically spanning $[-\delta q, +\delta q]$, where $\delta q \propto \sqrt{\hbar / m \omega}$ corresponds to the spread of the harmonic oscillator’s ground state wavefunction. We computed the Berry phase by parameterizing the spin-vibrational Hamiltonian (Eq.~\eqref{eq:vb4}) as $\hat{H}_{s-\text{vib}}[\eta_+, \eta_-]$, by replacing $\lambda_{k,4}^{q} \to \eta_+ \lambda_{k,4}^{q}$ for right-modes and $\lambda_{k,4}^{q} \to \eta_- \lambda_{k,4}^{q}$ for left-handed modes. The resulting coupling terms, $\eta_+ \lambda_{k,4}^{q} \hat{O}_k^q (J) \otimes \hat{L}_+^{\text{vib}}$ and $\eta_- \lambda_{k,4}^{q} \hat{O}_k^q (J) \otimes \hat{L}_-^{\text{vib}}$ couple the spin to the right- and left-handed vibrational modes, with strengths proportional to $\delta q$. These terms reflect the decomposition of angular momentum operators (e.g., related to $\hat{L}_y^{\text{vib}}$), capturing asymmetric chiral interactions. When $\eta_+ \neq \eta_-$, this asymmetry induces a non-trivial Berry phase, revealing the states’ chirality through the Berry curvature in $[\eta_+, \eta_-]$ space.

The eigenfunctions of the effective Hamiltonian $\hat{H}_{s-\text{vib}}$ were constructed in the circularly polarized basis:
\begin{align}\label{eq:eigen}
\ket{\psi^n [\eta_+,\eta_-]} = \sum_{m_J, n_\pm} c_{m_J, n_+, n_-} \ket{m_J, n_+, n_-},
\end{align} 
where $n_{\pm}$ labels the vibrational states, and the superscript $n$ represents the eigenstate index. Below, we discuss the single excitation for $n_{\pm} =1$, which can be easily generalized to $n_{\pm} \geq 1$. The eigenbasis for vibrational states were defined using $\ket{0}=\ket{0_+,0_-}$, $\ket{1_+}=\ket{1_+,0}$, $\ket{1_-}=\ket{0,1_-}$, with spin basis consisting of $m_{J}$. When both modes are excited, the basis transform as $\ket{2}=1/\sqrt{2}\left( \ket{2_+,0_-} - \ket{0_+,2_-} \right)$.  In the absence of SVC, the eigenstates can be represented in a decoupled basis as $\sum_{m_J,n_\pm} c_{m_J, n_+, n_-} \ket{m_J} \otimes (\ket{1_+} + \ket{1_-})$. The Zeeman term splits the spin states into $m_{+J}$ and $m_{-J}$, while the SVC term lifts the degeneracy of the vibrational states $\ket{1_+}$ and $\ket{1_-}$, resulting in the dressed spin-vibronic states $\ket{\psi^{1(2)}}$. Next, we discuss the geometric properties of dressed chiral vibronic states induced by the SVC terms in a spin-1 system.

\begin{figure}[h!]
\includegraphics[width=8.5cm,keepaspectratio]{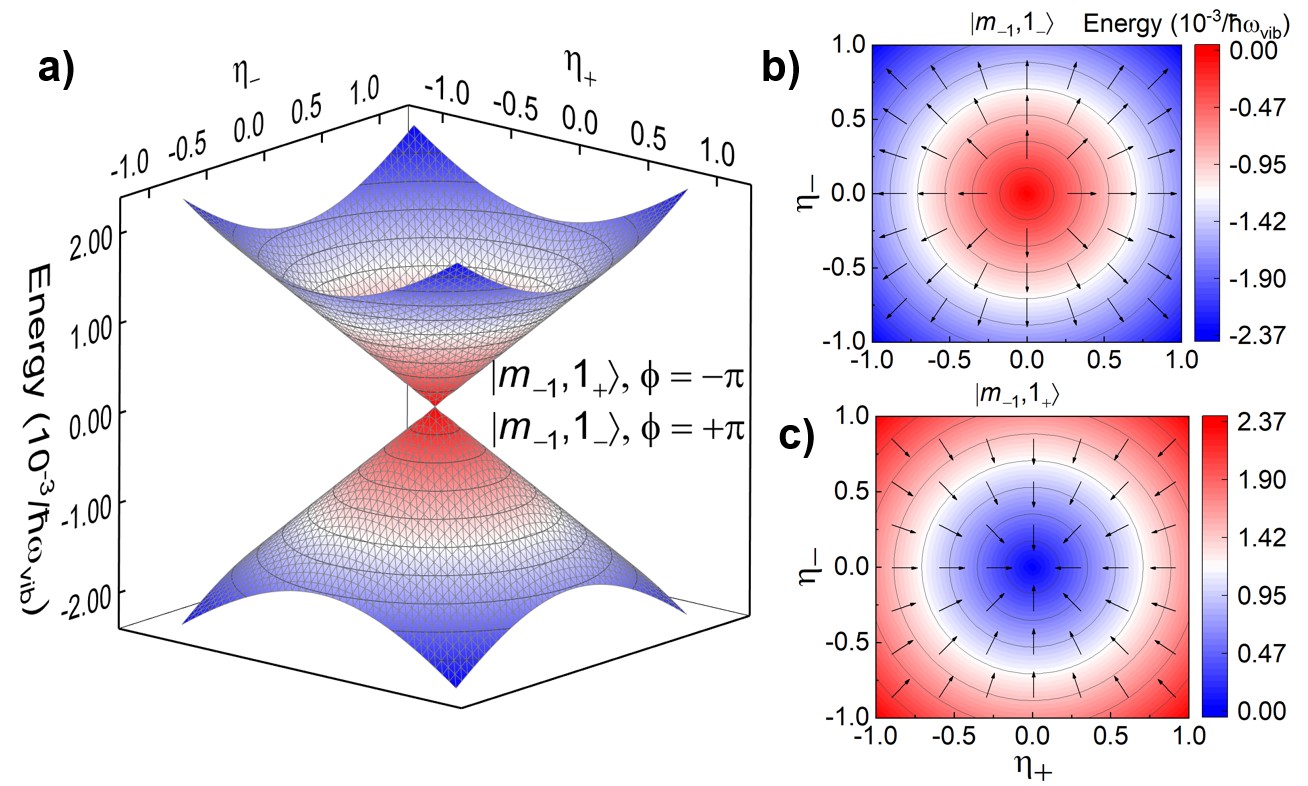}\\
\caption{ \small Topological character of spin-vibronic states and their vector field signatures.
\textbf{a)} spin-vibronic states $\ket{m_{-1},1_-}$ and $\ket{m_{-1},1_+}$ form a conical intersection in the  $[\eta_+, \eta_-]$ parameter space, acquiring Berry phases of $\pm \pi$ that reveal their nontrivial topology.
\textbf{b)} Outward-pointing vector field for state $\ket{m_{-1},1_-}$  and\textbf{ c)} inward-pointing vector field for state $\ket{m_{-1},1_+}$, exhibiting a dipole-like character.}
\label{fig:2}
\end{figure}

The effective Hamiltonian was diagonalized in the parameter space [-1,1], yielding eigenenergies and eigenfunctions, which were used to calculate the Berry flux and Berry phase (see SI for details). In Fig. \ref{fig:2}a, we show the eigenenergy surfaces for vibronic levels $\ket{m_{-1},1_-}$ and $\ket{m_{-1},1_+}$, which form a conical intersection and acquire Berry phase $+\pi$ and $-\pi$, respectively. In contrast, states such as $\ket{ m_{\pm J},0}$ and $\ket{m_{\pm J},2}$ do not acquire any Berry phase, confirming their trivial character. 

One can visualize the vector field $\hat{v}_{\mu}$in parameter space by calculating the expectation values of the operators $\hat{v}_x$ and $\hat{v}_y$, where $\hat{v}_x=\hat{J}_x\otimes \hat{L}_x^{vib}$, $\hat{v}_y=\hat{J}_y\otimes \hat{L}_y^{vib}$. The components of the vector field are given by $\langle \hat{v}_x \rangle=\bra{\psi^n[\eta_+,\eta_-]}\hat{v}_x\ket{\psi^n[\eta_+,\eta_-]}$ and 
$\langle \hat{v}_y \rangle=\bra{\psi^n[\eta_+,\eta_-]}\hat{v}_y\ket{\psi^n[\eta_+,\eta_-]}$. The direction of the vectors is determined by $\tan^{-1}{(\langle \hat{v}_y \rangle/\langle \hat{v}_y \rangle})$ and its magnitude is given by $\sqrt{\langle \hat{v}_x \rangle^2+\langle \hat{v}_y \rangle^2}$. The results (Fig. \ref{fig:2}b and c) reveal that the vector field for state $\ket{m_{-1},1_-}$  points outward away from the origin (conical intersection). In contrast, for state  $\ket{m_{-1},1_+}$, points towards the origin in parameter space. This is due to an effective monopole in parameter space, corresponding to $\pm\pi$-Berry phases. The combination of eigenenergy surfaces that form conical intersection and Berry phase demonstrates that $\ket{m_{-1},1_-}$ and $\ket{m_{-1},1_+}$ are spin-vibronic states with distinct geometric properties, a direct consequence of the SVC resulting form broken inversion and time-reversal symmetry. These results highlight the critical role of SVC in breaking degeneracies and inducing chirality in vibronic systems. 

\begin{table}[b]
  \centering
  \caption{Wave function composition of ground and first excited Kramer-doublet (KD).}
    \begin{tabular}{cr}
    \hline
    \multicolumn{1}{p{4em}}{ \centering KD (cm$^{-1}$)} & \multicolumn{1}{l}{\centering Wave-function} \\
    \hline
    0  &  73.6\%$\ket{\pm 1/2}$ + 26.4\%$\ket{\pm 5/2} $  \\
    634 &  90.8\%$\ket{\pm 3/2} $  + 9.2\%$\ket{\pm 3/2} $ \\
    990 &   73.6\%$\ket{\pm 5/2}$ + 26.4 \%$\ket{\pm 1/2}$\\
    \hline
    \end{tabular}%
  \label{tab:levels}%
\end{table}%

\begin{figure}[t]
\includegraphics[width=8.5cm,keepaspectratio]{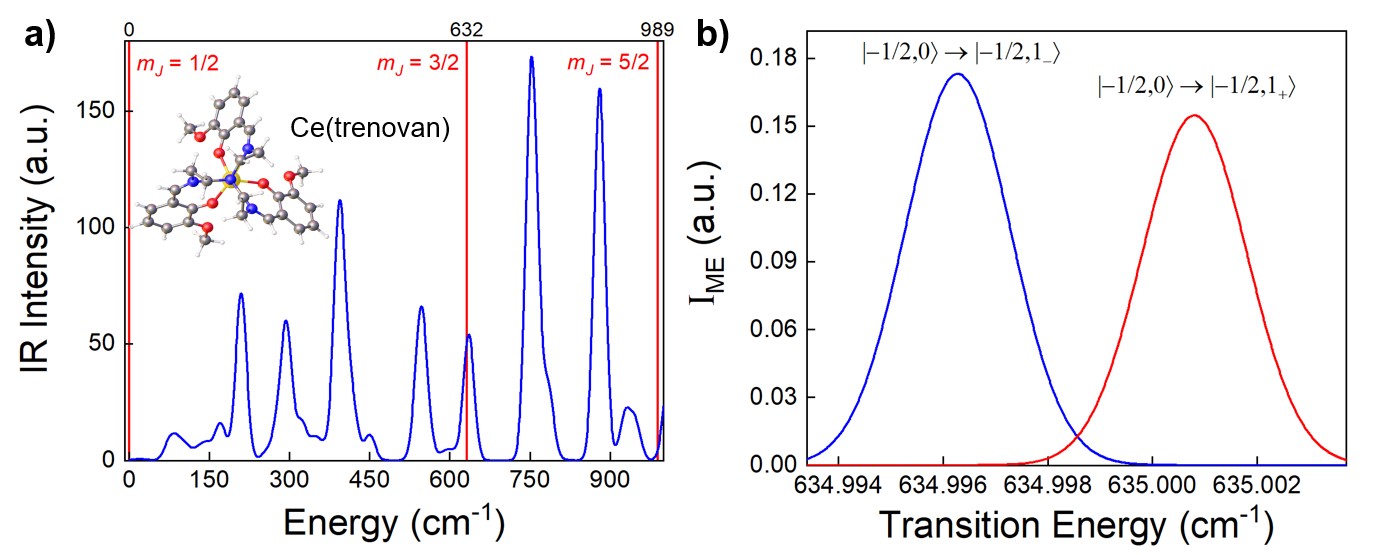}\\
\caption{\small Spin-vibronic states and selective excitation in Ce(trenovan).
a) Calculated vibrational spectrum (Gaussian-broadened, blue) with $f$-manifold crystal field levels $m_J$ (red vertical lines). The molecular structure of Ce(trenovan) is shown in the inset. b) SVC lifts the degeneracy of $\ket{-1/2,1_-}$ and $\ket{-1/2,1_+}$ states, each with angular momentum $\pm \hbar$, enabling selective electric- and magnetic-dipole transitions from $\ket{-1/2,0}$. The resulting transition energies and intensities are shown. }
\label{fig:3}
\end{figure}

\emph{Ce(trenovan) (molecular magnet)}.---We applied our framework to the molecular nanomagnet Ce(trenovan), trenovan = tris (3-methoxysalicylidene) amino) ethyl) amine, Fig. \ref{fig:3} a \cite{lucaccini2017electronic}, which features a trigonal symmetry of $C_3$. The molecular structure was fully optimized using density functional theory (DFT), and the far-infrared (IR) spectra were computed using the Gaussian software \cite{Gaussian}. Due to the $C_3$ symmetry, several vibrational modes transform according to the E-representation, making them ideal candidates for observing spin-vibronic effects.  The spin spectrum was obtained using the complete active space self-consistent field with spin-orbit coupling (CASSCF-SO) method, implemented in Molcas \cite{fdez2019openmolcas}. The presence of a large unquenched orbital angular momentum ($\hat{L}$) results in strong spin-orbit coupling, rendering the spin quantum number ($\hat{S}$) inadequate. Instead, the total angular momentum $\hat{\bf J}=\hat{\bf L}+\hat{\bf S}$ becomes the relevant quantum number. The ground multiplet ($^2F_{5/2}$) is separated from the first excited multiplet ($^2F_{7/2}$) by 1357 $cm^{-1}$, restricting the $^2F_{5/2}\to$$^2F_{7/2}$ magnetic transitions. Due to the crystal field, the $^2F_{5/2}$ multiplet further splits into $m_J$ sublevels. The wavefunction composition is provided in Table \ref{tab:levels}.

In the vibrational spectra, a degenerate mode at 635.30 $cm^{-1}$ lies close to the spin transition energy, as shown in Fig. \ref{fig:3}a. The Zeeman Hamiltonian, $\mu_B g_J B J_z$, was employed with $g_J=0.85$~\cite{abragam2012electron}, and $B=1$ Tesla to lift the degeneracy of the ground state $\ket{\pm 1/2}$.  The detailed methodology for computing the spin and vibrational spectra is provided in the SI.

To construct the SVC Hamiltonian, we distorted the equilibrium geometry along the normal mode vector of the E-mode and computed the crystal field levels at the CASSCF-SO level. The evolution of the crystal field parameters was fitted with second-order polynomials to extract the first- ($\partial B_k^q/\partial q_{a,b}$) and second-order  ($\partial^2 B_k^q/(\partial q_a\partial q_b)$) SVCs. These values are further used to determine $\lambda's$ for Eq. \ref{eq:vb4} in circular basis. Using the calculated SVC, we determined the Berry curvature and the Berry phase for the vibronic states. 
The Berry phase ($\phi$) acquired by  $\ket{-1/2,1_+}$ and  $\ket{-1/2,1_-}$ are $+\pi$ and $-\pi$ respectively. This confirms that $\ket{-1/2,1_-}$ and $\ket{-1/2,1_+}$ are spin-vibronic states with opposite handedness. 

The selection rules for transitions between spin-vibronic states are governed by the transition dipole matrix on the product basis $\hat{\mu}_{tot}=\hat{\mu}_{M}\otimes\left(\hat{\mu}_{E}^+\pm i\hat{\mu}_{E}^-\right)$, where $\hat{\mu}_{M}$ and $\hat{\mu}_{E}^{\pm}$ are magnetic and electric-dipole assisted transition operators.
The transition intensity ($I_{ME}$), which combines $\hat{\mu}_{M}$ and $\hat{\mu}_{E}^{\pm}$, is determined by the Einstein coefficients \cite{craig1998molecular,kragskow2022analysis}, $ B_{ME}=B_M\otimes B_E$  with $B_M=2\pi\mu_0/(3\hbar^2c^2)\hat{\mu}_{M}^\dagger\hat{\mu}_{M}$ and $B_E=2\pi/(3\hbar^2c^2\epsilon_0)(\hat{\mu}_{E}^\pm)^\dagger\hat{\mu}_{E}^\pm$, $\mu_0$ and $\epsilon_0$ are the permeability and permittivity of free space, respectively, $c$ is the speed of light. 
The allowed transitions,  the two distinct  peaks separated by $ \approx 0.005 cm^{-1}$ appears which corresponds to $\ket{-1/2,0}\to \ket{-1/2,1_+}$ and $\ket{-1/2,0}\to \ket{-1/2,1_-}$, Fig. \ref{fig:3}b. The separated transition energy is due to SVC couplings in the presence of an external Zeeman field, which lifts $\ket{-1/2}$. These results highlight the Ce(trenovan) potential for observing polarized peaks and provide a clear signature of spin-vibronic interactions, which can be resolved using high-resolution magneto-optical techniques, providing direct evidence of the spin-vibronic states \cite{rikken1997observation, yin2021chiral,jahnigen2023vibrational}.

This framework extends to systems with non-Abelian symmetry groups, summarized in Table~\ref{tab:symm}. High-symmetry groups, such as \(D_5\) and \(D_{5h}\), quench off-diagonal terms in the crystal field Hamiltonian, reducing decoherence from environmental noise and preserving the coherence of spin-vibronic states~\cite{liu2018symmetry}. Beyond these systems, our approach can be applied to transition metal-based molecular magnets, where crystal field effects dominate the spin-orbit coupling effects, inducing similar symmetry-breaking mechanisms~\cite{bistoni2021intrinsic}. These extensions highlight the versatility of our findings to other single molecular magnets.

\begin{table}[t]
  \centering
  \caption{Abelian and non-Abelian point symmetry groups with degenerate vibrational mode.}
    \begin{tabular}{c|p{9em}c}
    \hline
           \multicolumn{1}{c|}{Class} & \multicolumn{1}{p{9 em}}{\centering{Point symmetry group}}  & \multicolumn{1}{p{3em}}{} \\
    \hline
     Abelian &  $C_n$,  $C_{nh}$ & \textit{n} $\ge$ 3  \\
     \hline
     non-Abelian  & $C_{nv}$, $D_n$, $D_{nh}$    & \textit{n} $\ge$ 3 \\
          & $D_{nd}$    & \textit{n} $\ge$ 2  \\
           \hline
    \end{tabular}
  \label{tab:symm}
\end{table}

{\em Quantum Initialization}.---Our findings on selective excitation of spin-vibronic states open new avenues in chiral phononics and quantum optics~\cite{faure2018topological,zhang2024quantum}. In particular, the spin-vibronic dressed states exhibit distinct chirality characteristics critical to these applications. The ground state $\ket{m_J, 0}$, with zero Berry phase and no vibrational angular momentum, serves as a stable reference, while the excited states $\ket{m_J, 1_\mp}$, with Berry phases of $\pm\pi$ and vibrational angular momenta of $\mp 1$, enhance sensitivity and encoding capabilities. For quantum initialization, the contrast between the neutral ground state and the chiral excited states facilitates reliable state preparation. The $\pm\pi$ Berry phases, arising from conical intersections, ensure robustness, making these states ideal for quantum sensing and information encoding.  Applications include quantum sensing, where the sensitivity of spin-vibronic states to external fields enables precise measurements of rotational or chiral effects, and quantum information processing, where their geometric properties support fault-tolerant quantum gates. Additionally, initializing states with defined angular momentum offers robust protocols for quantum control~\cite{bistoni2021intrinsic}.


{\em Conclusion}.---Our results of chiral light-matter interactions in molecular and solid-state systems provide a foundation for future studies on geometric properties of spin-vibronic states in single molecular magnets\cite{gatteschi2006molecular}. The universality of the SVC  mechanism suggests its applicability across a wide range of materials, from molecular magnets to solid-state systems, offering new opportunities in chiral phononics, and quantum optics. Controlling spin-vibronic states through external fields or strain enhances their potential for applications in next-generation quantum technologies. Furthermore, these insights could impact quantum information processing by leveraging geometric phases for fault-tolerant quantum gates \cite{jones2000geometric, zanardi1999holonomic} and inform advances in chiral chemistry through a deeper understanding of molecular symmetry breaking \cite{barron1986symmetry, tang2010optical}.

\section{Acknowledgements}
The authors would like to acknowledge support from the U.S. Department of Energy, Office of Science, Office of Basic Energy Sciences, Established Program to Stimulate Competitive Research under Award Number DE-SC0022178.
\bibliographystyle{apsrev4-2}
\bibliography{ref}

\begin{thebibliography}{41}%
\makeatletter
\providecommand \@ifxundefined [1]{%
 \@ifx{#1\undefined}
}%
\providecommand \@ifnum [1]{%
 \ifnum #1\expandafter \@firstoftwo
 \else \expandafter \@secondoftwo
 \fi
}%
\providecommand \@ifx [1]{%
 \ifx #1\expandafter \@firstoftwo
 \else \expandafter \@secondoftwo
 \fi
}%
\providecommand \natexlab [1]{#1}%
\providecommand \enquote  [1]{``#1''}%
\providecommand \bibnamefont  [1]{#1}%
\providecommand \bibfnamefont [1]{#1}%
\providecommand \citenamefont [1]{#1}%
\providecommand \href@noop [0]{\@secondoftwo}%
\providecommand \href [0]{\begingroup \@sanitize@url \@href}%
\providecommand \@href[1]{\@@startlink{#1}\@@href}%
\providecommand \@@href[1]{\endgroup#1\@@endlink}%
\providecommand \@sanitize@url [0]{\catcode `\\12\catcode `\$12\catcode `\&12\catcode `\#12\catcode `\^12\catcode `\_12\catcode `\%12\relax}%
\providecommand \@@startlink[1]{}%
\providecommand \@@endlink[0]{}%
\providecommand \url  [0]{\begingroup\@sanitize@url \@url }%
\providecommand \@url [1]{\endgroup\@href {#1}{\urlprefix }}%
\providecommand \urlprefix  [0]{URL }%
\providecommand \Eprint [0]{\href }%
\providecommand \doibase [0]{https://doi.org/}%
\providecommand \selectlanguage [0]{\@gobble}%
\providecommand \bibinfo  [0]{\@secondoftwo}%
\providecommand \bibfield  [0]{\@secondoftwo}%
\providecommand \translation [1]{[#1]}%
\providecommand \BibitemOpen [0]{}%
\providecommand \bibitemStop [0]{}%
\providecommand \bibitemNoStop [0]{.\EOS\space}%
\providecommand \EOS [0]{\spacefactor3000\relax}%
\providecommand \BibitemShut  [1]{\csname bibitem#1\endcsname}%
\let\auto@bib@innerbib\@empty
\bibitem [{\citenamefont {Ueda}\ \emph {et~al.}(2023)\citenamefont {Ueda}, \citenamefont {Garc{\'\i}a-Fern{\'a}ndez}, \citenamefont {Agrestini}, \citenamefont {Romao}, \citenamefont {van~den Brink}, \citenamefont {Spaldin}, \citenamefont {Zhou},\ and\ \citenamefont {Staub}}]{ueda2023chiral}%
  \BibitemOpen
  \bibfield  {author} {\bibinfo {author} {\bibfnamefont {H.}~\bibnamefont {Ueda}}, \bibinfo {author} {\bibfnamefont {M.}~\bibnamefont {Garc{\'\i}a-Fern{\'a}ndez}}, \bibinfo {author} {\bibfnamefont {S.}~\bibnamefont {Agrestini}}, \bibinfo {author} {\bibfnamefont {C.~P.}\ \bibnamefont {Romao}}, \bibinfo {author} {\bibfnamefont {J.}~\bibnamefont {van~den Brink}}, \bibinfo {author} {\bibfnamefont {N.~A.}\ \bibnamefont {Spaldin}}, \bibinfo {author} {\bibfnamefont {K.-J.}\ \bibnamefont {Zhou}},\ and\ \bibinfo {author} {\bibfnamefont {U.}~\bibnamefont {Staub}},\ }\href@noop {} {\bibfield  {journal} {\bibinfo  {journal} {Nature}\ }\textbf {\bibinfo {volume} {618}},\ \bibinfo {pages} {946} (\bibinfo {year} {2023})}\BibitemShut {NoStop}%
\bibitem [{\citenamefont {Wang}\ \emph {et~al.}(2024)\citenamefont {Wang}, \citenamefont {Sun}, \citenamefont {Li},\ and\ \citenamefont {Zhang}}]{wang2024chiral}%
  \BibitemOpen
  \bibfield  {author} {\bibinfo {author} {\bibfnamefont {T.}~\bibnamefont {Wang}}, \bibinfo {author} {\bibfnamefont {H.}~\bibnamefont {Sun}}, \bibinfo {author} {\bibfnamefont {X.}~\bibnamefont {Li}},\ and\ \bibinfo {author} {\bibfnamefont {L.}~\bibnamefont {Zhang}},\ }\href@noop {} {\bibfield  {journal} {\bibinfo  {journal} {Nano Lett.}\ }\textbf {\bibinfo {volume} {24}},\ \bibinfo {pages} {4311} (\bibinfo {year} {2024})}\BibitemShut {NoStop}%
\bibitem [{\citenamefont {Aiello}\ \emph {et~al.}(2022)\citenamefont {Aiello}, \citenamefont {Abendroth}, \citenamefont {Abbas}, \citenamefont {Afanasev}, \citenamefont {Agarwal}, \citenamefont {Banerjee}, \citenamefont {Beratan}, \citenamefont {Belling}, \citenamefont {Berche}, \citenamefont {Botana} \emph {et~al.}}]{aiello2022chirality}%
  \BibitemOpen
  \bibfield  {author} {\bibinfo {author} {\bibfnamefont {C.~D.}\ \bibnamefont {Aiello}}, \bibinfo {author} {\bibfnamefont {J.~M.}\ \bibnamefont {Abendroth}}, \bibinfo {author} {\bibfnamefont {M.}~\bibnamefont {Abbas}}, \bibinfo {author} {\bibfnamefont {A.}~\bibnamefont {Afanasev}}, \bibinfo {author} {\bibfnamefont {S.}~\bibnamefont {Agarwal}}, \bibinfo {author} {\bibfnamefont {A.~S.}\ \bibnamefont {Banerjee}}, \bibinfo {author} {\bibfnamefont {D.~N.}\ \bibnamefont {Beratan}}, \bibinfo {author} {\bibfnamefont {J.~N.}\ \bibnamefont {Belling}}, \bibinfo {author} {\bibfnamefont {B.}~\bibnamefont {Berche}}, \bibinfo {author} {\bibfnamefont {A.}~\bibnamefont {Botana}}, \emph {et~al.},\ }\href@noop {} {\bibfield  {journal} {\bibinfo  {journal} {ACS Nano}\ }\textbf {\bibinfo {volume} {16}},\ \bibinfo {pages} {4989} (\bibinfo {year} {2022})}\BibitemShut {NoStop}%
\bibitem [{\citenamefont {Shushkov}(2024)}]{shushkov2024novel}%
  \BibitemOpen
  \bibfield  {author} {\bibinfo {author} {\bibfnamefont {P.}~\bibnamefont {Shushkov}},\ }\href@noop {} {\bibfield  {journal} {\bibinfo  {journal} {J. Chem. Phys.}\ }\textbf {\bibinfo {volume} {160}} (\bibinfo {year} {2024})}\BibitemShut {NoStop}%
\bibitem [{\citenamefont {Coh}(2023)}]{coh2023classification}%
  \BibitemOpen
  \bibfield  {author} {\bibinfo {author} {\bibfnamefont {S.}~\bibnamefont {Coh}},\ }\href@noop {} {\bibfield  {journal} {\bibinfo  {journal} {Phys. Rev. B}\ }\textbf {\bibinfo {volume} {108}},\ \bibinfo {pages} {134307} (\bibinfo {year} {2023})}\BibitemShut {NoStop}%
\bibitem [{\citenamefont {Lujan}\ \emph {et~al.}(2024)\citenamefont {Lujan}, \citenamefont {Choe}, \citenamefont {Chaudhary}, \citenamefont {Ye}, \citenamefont {Nnokwe}, \citenamefont {Rodriguez-Vega}, \citenamefont {He}, \citenamefont {Gao}, \citenamefont {Nunley}, \citenamefont {Baldini} \emph {et~al.}}]{lujan2024spin}%
  \BibitemOpen
  \bibfield  {author} {\bibinfo {author} {\bibfnamefont {D.}~\bibnamefont {Lujan}}, \bibinfo {author} {\bibfnamefont {J.}~\bibnamefont {Choe}}, \bibinfo {author} {\bibfnamefont {S.}~\bibnamefont {Chaudhary}}, \bibinfo {author} {\bibfnamefont {G.}~\bibnamefont {Ye}}, \bibinfo {author} {\bibfnamefont {C.}~\bibnamefont {Nnokwe}}, \bibinfo {author} {\bibfnamefont {M.}~\bibnamefont {Rodriguez-Vega}}, \bibinfo {author} {\bibfnamefont {J.}~\bibnamefont {He}}, \bibinfo {author} {\bibfnamefont {F.~Y.}\ \bibnamefont {Gao}}, \bibinfo {author} {\bibfnamefont {T.~N.}\ \bibnamefont {Nunley}}, \bibinfo {author} {\bibfnamefont {E.}~\bibnamefont {Baldini}}, \emph {et~al.},\ }\href@noop {} {\bibfield  {journal} {\bibinfo  {journal} {Proc. Natl. Acad. Sci.}\ }\textbf {\bibinfo {volume} {121}},\ \bibinfo {pages} {e2304360121} (\bibinfo {year} {2024})}\BibitemShut {NoStop}%
\bibitem [{\citenamefont {Chaudhary}\ \emph {et~al.}(2024)\citenamefont {Chaudhary}, \citenamefont {Juraschek}, \citenamefont {Rodriguez-Vega},\ and\ \citenamefont {Fiete}}]{chaudhary2024giant}%
  \BibitemOpen
  \bibfield  {author} {\bibinfo {author} {\bibfnamefont {S.}~\bibnamefont {Chaudhary}}, \bibinfo {author} {\bibfnamefont {D.~M.}\ \bibnamefont {Juraschek}}, \bibinfo {author} {\bibfnamefont {M.}~\bibnamefont {Rodriguez-Vega}},\ and\ \bibinfo {author} {\bibfnamefont {G.~A.}\ \bibnamefont {Fiete}},\ }\href@noop {} {\bibfield  {journal} {\bibinfo  {journal} {Phys. Rev. B}\ }\textbf {\bibinfo {volume} {110}},\ \bibinfo {pages} {094401} (\bibinfo {year} {2024})}\BibitemShut {NoStop}%
\bibitem [{\citenamefont {Chen}\ \emph {et~al.}(2018)\citenamefont {Chen}, \citenamefont {Zhang}, \citenamefont {Niu},\ and\ \citenamefont {Zhang}}]{chen2018chiral}%
  \BibitemOpen
  \bibfield  {author} {\bibinfo {author} {\bibfnamefont {H.}~\bibnamefont {Chen}}, \bibinfo {author} {\bibfnamefont {W.}~\bibnamefont {Zhang}}, \bibinfo {author} {\bibfnamefont {Q.}~\bibnamefont {Niu}},\ and\ \bibinfo {author} {\bibfnamefont {L.}~\bibnamefont {Zhang}},\ }\href@noop {} {\bibfield  {journal} {\bibinfo  {journal} {2D Materials}\ }\textbf {\bibinfo {volume} {6}},\ \bibinfo {pages} {012002} (\bibinfo {year} {2018})}\BibitemShut {NoStop}%
\bibitem [{\citenamefont {Zhang}\ and\ \citenamefont {Niu}(2015)}]{zhang2015chiral}%
  \BibitemOpen
  \bibfield  {author} {\bibinfo {author} {\bibfnamefont {L.}~\bibnamefont {Zhang}}\ and\ \bibinfo {author} {\bibfnamefont {Q.}~\bibnamefont {Niu}},\ }\href@noop {} {\bibfield  {journal} {\bibinfo  {journal} {Phys. Rev. Lett.}\ }\textbf {\bibinfo {volume} {115}},\ \bibinfo {pages} {115502} (\bibinfo {year} {2015})}\BibitemShut {NoStop}%
\bibitem [{\citenamefont {Zhang}\ and\ \citenamefont {Niu}(2014)}]{zhang2014angular}%
  \BibitemOpen
  \bibfield  {author} {\bibinfo {author} {\bibfnamefont {L.}~\bibnamefont {Zhang}}\ and\ \bibinfo {author} {\bibfnamefont {Q.}~\bibnamefont {Niu}},\ }\href@noop {} {\bibfield  {journal} {\bibinfo  {journal} {Phys. Rev. Lett.}\ }\textbf {\bibinfo {volume} {112}},\ \bibinfo {pages} {085503} (\bibinfo {year} {2014})}\BibitemShut {NoStop}%
\bibitem [{\citenamefont {Hernandez}\ \emph {et~al.}(2023)\citenamefont {Hernandez}, \citenamefont {Baydin}, \citenamefont {Chaudhary}, \citenamefont {Tay}, \citenamefont {Katayama}, \citenamefont {Takeda}, \citenamefont {Nojiri}, \citenamefont {Okazaki}, \citenamefont {Rappl}, \citenamefont {Abramof} \emph {et~al.}}]{hernandez2023observation}%
  \BibitemOpen
  \bibfield  {author} {\bibinfo {author} {\bibfnamefont {F.~G.}\ \bibnamefont {Hernandez}}, \bibinfo {author} {\bibfnamefont {A.}~\bibnamefont {Baydin}}, \bibinfo {author} {\bibfnamefont {S.}~\bibnamefont {Chaudhary}}, \bibinfo {author} {\bibfnamefont {F.}~\bibnamefont {Tay}}, \bibinfo {author} {\bibfnamefont {I.}~\bibnamefont {Katayama}}, \bibinfo {author} {\bibfnamefont {J.}~\bibnamefont {Takeda}}, \bibinfo {author} {\bibfnamefont {H.}~\bibnamefont {Nojiri}}, \bibinfo {author} {\bibfnamefont {A.~K.}\ \bibnamefont {Okazaki}}, \bibinfo {author} {\bibfnamefont {P.~H.}\ \bibnamefont {Rappl}}, \bibinfo {author} {\bibfnamefont {E.}~\bibnamefont {Abramof}}, \emph {et~al.},\ }\href@noop {} {\bibfield  {journal} {\bibinfo  {journal} {Sci. Adv.}\ }\textbf {\bibinfo {volume} {9}},\ \bibinfo {pages} {eadj4074} (\bibinfo {year} {2023})}\BibitemShut {NoStop}%
\bibitem [{\citenamefont {Berry}(1984)}]{berry1984quantal}%
  \BibitemOpen
  \bibfield  {author} {\bibinfo {author} {\bibfnamefont {M.~V.}\ \bibnamefont {Berry}},\ }\href@noop {} {\bibfield  {journal} {\bibinfo  {journal} {Proceedings of the Royal Society of London. A. Mathematical and Physical Sciences}\ }\textbf {\bibinfo {volume} {392}},\ \bibinfo {pages} {45} (\bibinfo {year} {1984})}\BibitemShut {NoStop}%
\bibitem [{\citenamefont {Yarkony}(1996)}]{yarkony1996diabolical}%
  \BibitemOpen
  \bibfield  {author} {\bibinfo {author} {\bibfnamefont {D.~R.}\ \bibnamefont {Yarkony}},\ }\href@noop {} {\bibfield  {journal} {\bibinfo  {journal} {Rev. Mod. Phys.}\ }\textbf {\bibinfo {volume} {68}},\ \bibinfo {pages} {985} (\bibinfo {year} {1996})}\BibitemShut {NoStop}%
\bibitem [{\citenamefont {Canali}\ \emph {et~al.}(2003)\citenamefont {Canali}, \citenamefont {Cehovin},\ and\ \citenamefont {MacDonald}}]{canali2003chern}%
  \BibitemOpen
  \bibfield  {author} {\bibinfo {author} {\bibfnamefont {C.~M.}\ \bibnamefont {Canali}}, \bibinfo {author} {\bibfnamefont {A.}~\bibnamefont {Cehovin}},\ and\ \bibinfo {author} {\bibfnamefont {A.}~\bibnamefont {MacDonald}},\ }\href@noop {} {\bibfield  {journal} {\bibinfo  {journal} {Phys. Rev. Lett.}\ }\textbf {\bibinfo {volume} {91}},\ \bibinfo {pages} {046805} (\bibinfo {year} {2003})}\BibitemShut {NoStop}%
\bibitem [{\citenamefont {Valahu}\ \emph {et~al.}(2023)\citenamefont {Valahu}, \citenamefont {Olaya-Agudelo}, \citenamefont {MacDonell}, \citenamefont {Navickas}, \citenamefont {Rao}, \citenamefont {Millican}, \citenamefont {P{\'e}rez-S{\'a}nchez}, \citenamefont {Yuen-Zhou}, \citenamefont {Biercuk}, \citenamefont {Hempel} \emph {et~al.}}]{valahu2023direct}%
  \BibitemOpen
  \bibfield  {author} {\bibinfo {author} {\bibfnamefont {C.~H.}\ \bibnamefont {Valahu}}, \bibinfo {author} {\bibfnamefont {V.~C.}\ \bibnamefont {Olaya-Agudelo}}, \bibinfo {author} {\bibfnamefont {R.~J.}\ \bibnamefont {MacDonell}}, \bibinfo {author} {\bibfnamefont {T.}~\bibnamefont {Navickas}}, \bibinfo {author} {\bibfnamefont {A.~D.}\ \bibnamefont {Rao}}, \bibinfo {author} {\bibfnamefont {M.~J.}\ \bibnamefont {Millican}}, \bibinfo {author} {\bibfnamefont {J.~B.}\ \bibnamefont {P{\'e}rez-S{\'a}nchez}}, \bibinfo {author} {\bibfnamefont {J.}~\bibnamefont {Yuen-Zhou}}, \bibinfo {author} {\bibfnamefont {M.~J.}\ \bibnamefont {Biercuk}}, \bibinfo {author} {\bibfnamefont {C.}~\bibnamefont {Hempel}}, \emph {et~al.},\ }\href@noop {} {\bibfield  {journal} {\bibinfo  {journal} {Nat. Chem.}\ }\textbf {\bibinfo {volume} {15}},\ \bibinfo {pages} {1503} (\bibinfo {year} {2023})}\BibitemShut {NoStop}%
\bibitem [{\citenamefont {Lucaccini}\ \emph {et~al.}(2017)\citenamefont {Lucaccini}, \citenamefont {Baldov{\'\i}}, \citenamefont {Chelazzi}, \citenamefont {Barra}, \citenamefont {Grepioni}, \citenamefont {Costes},\ and\ \citenamefont {Sorace}}]{lucaccini2017electronic}%
  \BibitemOpen
  \bibfield  {author} {\bibinfo {author} {\bibfnamefont {E.}~\bibnamefont {Lucaccini}}, \bibinfo {author} {\bibfnamefont {J.~J.}\ \bibnamefont {Baldov{\'\i}}}, \bibinfo {author} {\bibfnamefont {L.}~\bibnamefont {Chelazzi}}, \bibinfo {author} {\bibfnamefont {A.-L.}\ \bibnamefont {Barra}}, \bibinfo {author} {\bibfnamefont {F.}~\bibnamefont {Grepioni}}, \bibinfo {author} {\bibfnamefont {J.-P.}\ \bibnamefont {Costes}},\ and\ \bibinfo {author} {\bibfnamefont {L.}~\bibnamefont {Sorace}},\ }\href@noop {} {\bibfield  {journal} {\bibinfo  {journal} {Inorg. Chem.}\ }\textbf {\bibinfo {volume} {56}},\ \bibinfo {pages} {4728} (\bibinfo {year} {2017})}\BibitemShut {NoStop}%
\bibitem [{\citenamefont {Dergachev}\ \emph {et~al.}(2025)\citenamefont {Dergachev}, \citenamefont {Chibotaru},\ and\ \citenamefont {Varganov}}]{dergachev2025ab}%
  \BibitemOpen
  \bibfield  {author} {\bibinfo {author} {\bibfnamefont {V.~D.}\ \bibnamefont {Dergachev}}, \bibinfo {author} {\bibfnamefont {L.~F.}\ \bibnamefont {Chibotaru}},\ and\ \bibinfo {author} {\bibfnamefont {S.~A.}\ \bibnamefont {Varganov}},\ }\href@noop {} {\bibfield  {journal} {\bibinfo  {journal} {J. Phys. Chem. Lett.}\ }\textbf {\bibinfo {volume} {16}},\ \bibinfo {pages} {2309} (\bibinfo {year} {2025})}\BibitemShut {NoStop}%
\bibitem [{\citenamefont {Qi}\ and\ \citenamefont {Zhang}(2011)}]{qi2011topological}%
  \BibitemOpen
  \bibfield  {author} {\bibinfo {author} {\bibfnamefont {X.-L.}\ \bibnamefont {Qi}}\ and\ \bibinfo {author} {\bibfnamefont {S.-C.}\ \bibnamefont {Zhang}},\ }\href@noop {} {\bibfield  {journal} {\bibinfo  {journal} {Rev. Mod. Phys.}\ }\textbf {\bibinfo {volume} {83}},\ \bibinfo {pages} {1057} (\bibinfo {year} {2011})}\BibitemShut {NoStop}%
\bibitem [{\citenamefont {Ungur}\ and\ \citenamefont {Chibotaru}(2017)}]{ungur2017ab}%
  \BibitemOpen
  \bibfield  {author} {\bibinfo {author} {\bibfnamefont {L.}~\bibnamefont {Ungur}}\ and\ \bibinfo {author} {\bibfnamefont {L.~F.}\ \bibnamefont {Chibotaru}},\ }\href@noop {} {\bibfield  {journal} {\bibinfo  {journal} {Eur. J. Chem.}\ }\textbf {\bibinfo {volume} {23}},\ \bibinfo {pages} {3708} (\bibinfo {year} {2017})}\BibitemShut {NoStop}%
\bibitem [{\citenamefont {Ji}\ \emph {et~al.}(2025)\citenamefont {Ji}, \citenamefont {Zhao}, \citenamefont {Li},\ and\ \citenamefont {Chen}}]{ji2025circularly}%
  \BibitemOpen
  \bibfield  {author} {\bibinfo {author} {\bibfnamefont {M.-J.}\ \bibnamefont {Ji}}, \bibinfo {author} {\bibfnamefont {W.-L.}\ \bibnamefont {Zhao}}, \bibinfo {author} {\bibfnamefont {M.}~\bibnamefont {Li}},\ and\ \bibinfo {author} {\bibfnamefont {C.-F.}\ \bibnamefont {Chen}},\ }\href@noop {} {\bibfield  {journal} {\bibinfo  {journal} {Nat. Commun.}\ }\textbf {\bibinfo {volume} {16}},\ \bibinfo {pages} {2940} (\bibinfo {year} {2025})}\BibitemShut {NoStop}%
\bibitem [{\citenamefont {Elmers}\ \emph {et~al.}(2025)\citenamefont {Elmers}, \citenamefont {Tkach}, \citenamefont {Lytvynenko}, \citenamefont {Yogi}, \citenamefont {Schmitt}, \citenamefont {Biswas}, \citenamefont {Liu}, \citenamefont {Chernov}, \citenamefont {Nguyen}, \citenamefont {Hoesch} \emph {et~al.}}]{elmers2025chirality}%
  \BibitemOpen
  \bibfield  {author} {\bibinfo {author} {\bibfnamefont {H.}~\bibnamefont {Elmers}}, \bibinfo {author} {\bibfnamefont {O.}~\bibnamefont {Tkach}}, \bibinfo {author} {\bibfnamefont {Y.}~\bibnamefont {Lytvynenko}}, \bibinfo {author} {\bibfnamefont {P.}~\bibnamefont {Yogi}}, \bibinfo {author} {\bibfnamefont {M.}~\bibnamefont {Schmitt}}, \bibinfo {author} {\bibfnamefont {D.}~\bibnamefont {Biswas}}, \bibinfo {author} {\bibfnamefont {J.}~\bibnamefont {Liu}}, \bibinfo {author} {\bibfnamefont {S.}~\bibnamefont {Chernov}}, \bibinfo {author} {\bibfnamefont {Q.}~\bibnamefont {Nguyen}}, \bibinfo {author} {\bibfnamefont {M.}~\bibnamefont {Hoesch}}, \emph {et~al.},\ }\href@noop {} {\bibfield  {journal} {\bibinfo  {journal} {Phys. Rev. Lett.}\ }\textbf {\bibinfo {volume} {134}},\ \bibinfo {pages} {096401} (\bibinfo {year} {2025})}\BibitemShut {NoStop}%
\bibitem [{\citenamefont {Tikhonov}\ \emph {et~al.}(2022)\citenamefont {Tikhonov}, \citenamefont {Blech}, \citenamefont {Leibscher}, \citenamefont {Greenman}, \citenamefont {Schnell},\ and\ \citenamefont {Koch}}]{tikhonov2022pump}%
  \BibitemOpen
  \bibfield  {author} {\bibinfo {author} {\bibfnamefont {D.~S.}\ \bibnamefont {Tikhonov}}, \bibinfo {author} {\bibfnamefont {A.}~\bibnamefont {Blech}}, \bibinfo {author} {\bibfnamefont {M.}~\bibnamefont {Leibscher}}, \bibinfo {author} {\bibfnamefont {L.}~\bibnamefont {Greenman}}, \bibinfo {author} {\bibfnamefont {M.}~\bibnamefont {Schnell}},\ and\ \bibinfo {author} {\bibfnamefont {C.~P.}\ \bibnamefont {Koch}},\ }\href@noop {} {\bibfield  {journal} {\bibinfo  {journal} {Sci. Adv.}\ }\textbf {\bibinfo {volume} {8}},\ \bibinfo {pages} {eade0311} (\bibinfo {year} {2022})}\BibitemShut {NoStop}%
\bibitem [{\citenamefont {Huang}\ \emph {et~al.}(2023)\citenamefont {Huang}, \citenamefont {Sun}, \citenamefont {Wu}, \citenamefont {Liu}, \citenamefont {Zhan}, \citenamefont {Wang},\ and\ \citenamefont {Gao}}]{huang2023circularly}%
  \BibitemOpen
  \bibfield  {author} {\bibinfo {author} {\bibfnamefont {H.}~\bibnamefont {Huang}}, \bibinfo {author} {\bibfnamefont {R.}~\bibnamefont {Sun}}, \bibinfo {author} {\bibfnamefont {X.-F.}\ \bibnamefont {Wu}}, \bibinfo {author} {\bibfnamefont {Y.}~\bibnamefont {Liu}}, \bibinfo {author} {\bibfnamefont {J.-Z.}\ \bibnamefont {Zhan}}, \bibinfo {author} {\bibfnamefont {B.-W.}\ \bibnamefont {Wang}},\ and\ \bibinfo {author} {\bibfnamefont {S.}~\bibnamefont {Gao}},\ }\href@noop {} {\bibfield  {journal} {\bibinfo  {journal} {Dalton Trans.}\ }\textbf {\bibinfo {volume} {52}},\ \bibinfo {pages} {7646} (\bibinfo {year} {2023})}\BibitemShut {NoStop}%
\bibitem [{\citenamefont {Baydin}\ \emph {et~al.}(2022)\citenamefont {Baydin}, \citenamefont {Hernandez}, \citenamefont {Rodriguez-Vega}, \citenamefont {Okazaki}, \citenamefont {Tay}, \citenamefont {Noe}, \citenamefont {Katayama}, \citenamefont {Takeda}, \citenamefont {Nojiri}, \citenamefont {Rappl} \emph {et~al.}}]{baydin2022magnetic}%
  \BibitemOpen
  \bibfield  {author} {\bibinfo {author} {\bibfnamefont {A.}~\bibnamefont {Baydin}}, \bibinfo {author} {\bibfnamefont {F.~G.}\ \bibnamefont {Hernandez}}, \bibinfo {author} {\bibfnamefont {M.}~\bibnamefont {Rodriguez-Vega}}, \bibinfo {author} {\bibfnamefont {A.~K.}\ \bibnamefont {Okazaki}}, \bibinfo {author} {\bibfnamefont {F.}~\bibnamefont {Tay}}, \bibinfo {author} {\bibfnamefont {G.~T.}\ \bibnamefont {Noe}}, \bibinfo {author} {\bibfnamefont {I.}~\bibnamefont {Katayama}}, \bibinfo {author} {\bibfnamefont {J.}~\bibnamefont {Takeda}}, \bibinfo {author} {\bibfnamefont {H.}~\bibnamefont {Nojiri}}, \bibinfo {author} {\bibfnamefont {P.~H.}\ \bibnamefont {Rappl}}, \emph {et~al.},\ }\href@noop {} {\bibfield  {journal} {\bibinfo  {journal} {Phys. Rev. Lett.}\ }\textbf {\bibinfo {volume} {128}},\ \bibinfo {pages} {075901} (\bibinfo {year} {2022})}\BibitemShut {NoStop}%
\bibitem [{\citenamefont {Frisch}\ \emph {et~al.}(2016)\citenamefont {Frisch}, \citenamefont {Trucks}, \citenamefont {Schlegel}, \citenamefont {Scuseria}, \citenamefont {Robb}, \citenamefont {Cheeseman}, \citenamefont {Scalmani}, \citenamefont {Barone}, \citenamefont {Petersson}, \citenamefont {Nakatsuji} \emph {et~al.}}]{Gaussian}%
  \BibitemOpen
  \bibfield  {author} {\bibinfo {author} {\bibfnamefont {M.}~\bibnamefont {Frisch}}, \bibinfo {author} {\bibfnamefont {G.}~\bibnamefont {Trucks}}, \bibinfo {author} {\bibfnamefont {H.}~\bibnamefont {Schlegel}}, \bibinfo {author} {\bibfnamefont {G.}~\bibnamefont {Scuseria}}, \bibinfo {author} {\bibfnamefont {M.}~\bibnamefont {Robb}}, \bibinfo {author} {\bibfnamefont {J.}~\bibnamefont {Cheeseman}}, \bibinfo {author} {\bibfnamefont {G.}~\bibnamefont {Scalmani}}, \bibinfo {author} {\bibfnamefont {V.}~\bibnamefont {Barone}}, \bibinfo {author} {\bibfnamefont {G.}~\bibnamefont {Petersson}}, \bibinfo {author} {\bibfnamefont {H.}~\bibnamefont {Nakatsuji}}, \emph {et~al.},\ }\href@noop {} {\bibfield  {journal} {\bibinfo  {journal} {Gaussian Inc. Wallingford CT}\ } (\bibinfo {year} {2016})}\BibitemShut {NoStop}%
\bibitem [{\citenamefont {Fdez.~Galvan}\ \emph {et~al.}(2019)\citenamefont {Fdez.~Galvan}, \citenamefont {Vacher}, \citenamefont {Alavi}, \citenamefont {Angeli}, \citenamefont {Aquilante}, \citenamefont {Autschbach}, \citenamefont {Bao}, \citenamefont {Bokarev}, \citenamefont {Bogdanov}, \citenamefont {Carlson} \emph {et~al.}}]{fdez2019openmolcas}%
  \BibitemOpen
  \bibfield  {author} {\bibinfo {author} {\bibfnamefont {I.}~\bibnamefont {Fdez.~Galvan}}, \bibinfo {author} {\bibfnamefont {M.}~\bibnamefont {Vacher}}, \bibinfo {author} {\bibfnamefont {A.}~\bibnamefont {Alavi}}, \bibinfo {author} {\bibfnamefont {C.}~\bibnamefont {Angeli}}, \bibinfo {author} {\bibfnamefont {F.}~\bibnamefont {Aquilante}}, \bibinfo {author} {\bibfnamefont {J.}~\bibnamefont {Autschbach}}, \bibinfo {author} {\bibfnamefont {J.~J.}\ \bibnamefont {Bao}}, \bibinfo {author} {\bibfnamefont {S.~I.}\ \bibnamefont {Bokarev}}, \bibinfo {author} {\bibfnamefont {N.~A.}\ \bibnamefont {Bogdanov}}, \bibinfo {author} {\bibfnamefont {R.~K.}\ \bibnamefont {Carlson}}, \emph {et~al.},\ }\href@noop {} {\bibfield  {journal} {\bibinfo  {journal} {J. Chem. Theory Comput.}\ }\textbf {\bibinfo {volume} {15}},\ \bibinfo {pages} {5925} (\bibinfo {year} {2019})}\BibitemShut {NoStop}%
\bibitem [{\citenamefont {Abragam}\ and\ \citenamefont {Bleaney}(2012)}]{abragam2012electron}%
  \BibitemOpen
  \bibfield  {author} {\bibinfo {author} {\bibfnamefont {A.}~\bibnamefont {Abragam}}\ and\ \bibinfo {author} {\bibfnamefont {B.}~\bibnamefont {Bleaney}},\ }\href@noop {} {\emph {\bibinfo {title} {Electron paramagnetic resonance of transition ions}}}\ (\bibinfo  {publisher} {OUP oxford},\ \bibinfo {year} {2012})\BibitemShut {NoStop}%
\bibitem [{\citenamefont {Craig}\ and\ \citenamefont {Thirunamachandran}(1998)}]{craig1998molecular}%
  \BibitemOpen
  \bibfield  {author} {\bibinfo {author} {\bibfnamefont {D.~P.}\ \bibnamefont {Craig}}\ and\ \bibinfo {author} {\bibfnamefont {T.}~\bibnamefont {Thirunamachandran}},\ }\href@noop {} {\emph {\bibinfo {title} {Molecular quantum electrodynamics}}}\ (\bibinfo  {publisher} {Dover Publications},\ \bibinfo {year} {1998})\BibitemShut {NoStop}%
\bibitem [{\citenamefont {Kragskow}\ \emph {et~al.}(2022)\citenamefont {Kragskow}, \citenamefont {Marbey}, \citenamefont {Buch}, \citenamefont {Nehrkorn}, \citenamefont {Ozerov}, \citenamefont {Piligkos}, \citenamefont {Hill},\ and\ \citenamefont {Chilton}}]{kragskow2022analysis}%
  \BibitemOpen
  \bibfield  {author} {\bibinfo {author} {\bibfnamefont {J.~G.}\ \bibnamefont {Kragskow}}, \bibinfo {author} {\bibfnamefont {J.}~\bibnamefont {Marbey}}, \bibinfo {author} {\bibfnamefont {C.~D.}\ \bibnamefont {Buch}}, \bibinfo {author} {\bibfnamefont {J.}~\bibnamefont {Nehrkorn}}, \bibinfo {author} {\bibfnamefont {M.}~\bibnamefont {Ozerov}}, \bibinfo {author} {\bibfnamefont {S.}~\bibnamefont {Piligkos}}, \bibinfo {author} {\bibfnamefont {S.}~\bibnamefont {Hill}},\ and\ \bibinfo {author} {\bibfnamefont {N.~F.}\ \bibnamefont {Chilton}},\ }\href@noop {} {\bibfield  {journal} {\bibinfo  {journal} {Nat. Commun.}\ }\textbf {\bibinfo {volume} {13}},\ \bibinfo {pages} {825} (\bibinfo {year} {2022})}\BibitemShut {NoStop}%
\bibitem [{\citenamefont {Rikken}\ and\ \citenamefont {Raupach}(1997)}]{rikken1997observation}%
  \BibitemOpen
  \bibfield  {author} {\bibinfo {author} {\bibfnamefont {G.}~\bibnamefont {Rikken}}\ and\ \bibinfo {author} {\bibfnamefont {E.}~\bibnamefont {Raupach}},\ }\href@noop {} {\bibfield  {journal} {\bibinfo  {journal} {Nature}\ }\textbf {\bibinfo {volume} {390}},\ \bibinfo {pages} {493} (\bibinfo {year} {1997})}\BibitemShut {NoStop}%
\bibitem [{\citenamefont {Yin}\ \emph {et~al.}(2021)\citenamefont {Yin}, \citenamefont {Ulman}, \citenamefont {Liu}, \citenamefont {Granados~del {\'A}guila}, \citenamefont {Huang}, \citenamefont {Zhang}, \citenamefont {Serra}, \citenamefont {Sedmidubsky}, \citenamefont {Sofer}, \citenamefont {Quek} \emph {et~al.}}]{yin2021chiral}%
  \BibitemOpen
  \bibfield  {author} {\bibinfo {author} {\bibfnamefont {T.}~\bibnamefont {Yin}}, \bibinfo {author} {\bibfnamefont {K.~A.}\ \bibnamefont {Ulman}}, \bibinfo {author} {\bibfnamefont {S.}~\bibnamefont {Liu}}, \bibinfo {author} {\bibfnamefont {A.}~\bibnamefont {Granados~del {\'A}guila}}, \bibinfo {author} {\bibfnamefont {Y.}~\bibnamefont {Huang}}, \bibinfo {author} {\bibfnamefont {L.}~\bibnamefont {Zhang}}, \bibinfo {author} {\bibfnamefont {M.}~\bibnamefont {Serra}}, \bibinfo {author} {\bibfnamefont {D.}~\bibnamefont {Sedmidubsky}}, \bibinfo {author} {\bibfnamefont {Z.}~\bibnamefont {Sofer}}, \bibinfo {author} {\bibfnamefont {S.~Y.}\ \bibnamefont {Quek}}, \emph {et~al.},\ }\href@noop {} {\bibfield  {journal} {\bibinfo  {journal} {Adv. Mater.}\ }\textbf {\bibinfo {volume} {33}},\ \bibinfo {pages} {2101618} (\bibinfo {year} {2021})}\BibitemShut {NoStop}%
\bibitem [{\citenamefont {J{\"a}hnigen}(2023)}]{jahnigen2023vibrational}%
  \BibitemOpen
  \bibfield  {author} {\bibinfo {author} {\bibfnamefont {S.}~\bibnamefont {J{\"a}hnigen}},\ }\href@noop {} {\bibfield  {journal} {\bibinfo  {journal} {Angew. Chem., Int. Ed.}\ }\textbf {\bibinfo {volume} {62}},\ \bibinfo {pages} {e202303595} (\bibinfo {year} {2023})}\BibitemShut {NoStop}%
\bibitem [{\citenamefont {Liu}\ \emph {et~al.}(2018)\citenamefont {Liu}, \citenamefont {Chen},\ and\ \citenamefont {Tong}}]{liu2018symmetry}%
  \BibitemOpen
  \bibfield  {author} {\bibinfo {author} {\bibfnamefont {J.-L.}\ \bibnamefont {Liu}}, \bibinfo {author} {\bibfnamefont {Y.-C.}\ \bibnamefont {Chen}},\ and\ \bibinfo {author} {\bibfnamefont {M.-L.}\ \bibnamefont {Tong}},\ }\href@noop {} {\bibfield  {journal} {\bibinfo  {journal} {Chem. Soc. Rev.}\ }\textbf {\bibinfo {volume} {47}},\ \bibinfo {pages} {2431} (\bibinfo {year} {2018})}\BibitemShut {NoStop}%
\bibitem [{\citenamefont {Bistoni}\ \emph {et~al.}(2021)\citenamefont {Bistoni}, \citenamefont {Mauri},\ and\ \citenamefont {Calandra}}]{bistoni2021intrinsic}%
  \BibitemOpen
  \bibfield  {author} {\bibinfo {author} {\bibfnamefont {O.}~\bibnamefont {Bistoni}}, \bibinfo {author} {\bibfnamefont {F.}~\bibnamefont {Mauri}},\ and\ \bibinfo {author} {\bibfnamefont {M.}~\bibnamefont {Calandra}},\ }\href@noop {} {\bibfield  {journal} {\bibinfo  {journal} {Phys. Rev. Lett.}\ }\textbf {\bibinfo {volume} {126}},\ \bibinfo {pages} {225703} (\bibinfo {year} {2021})}\BibitemShut {NoStop}%
\bibitem [{\citenamefont {Faure}\ \emph {et~al.}(2018)\citenamefont {Faure}, \citenamefont {Takayoshi}, \citenamefont {Petit}, \citenamefont {Simonet}, \citenamefont {Raymond}, \citenamefont {Regnault}, \citenamefont {Boehm}, \citenamefont {White}, \citenamefont {M{\aa}nsson}, \citenamefont {R{\"u}egg} \emph {et~al.}}]{faure2018topological}%
  \BibitemOpen
  \bibfield  {author} {\bibinfo {author} {\bibfnamefont {Q.}~\bibnamefont {Faure}}, \bibinfo {author} {\bibfnamefont {S.}~\bibnamefont {Takayoshi}}, \bibinfo {author} {\bibfnamefont {S.}~\bibnamefont {Petit}}, \bibinfo {author} {\bibfnamefont {V.}~\bibnamefont {Simonet}}, \bibinfo {author} {\bibfnamefont {S.}~\bibnamefont {Raymond}}, \bibinfo {author} {\bibfnamefont {L.-P.}\ \bibnamefont {Regnault}}, \bibinfo {author} {\bibfnamefont {M.}~\bibnamefont {Boehm}}, \bibinfo {author} {\bibfnamefont {J.~S.}\ \bibnamefont {White}}, \bibinfo {author} {\bibfnamefont {M.}~\bibnamefont {M{\aa}nsson}}, \bibinfo {author} {\bibfnamefont {C.}~\bibnamefont {R{\"u}egg}}, \emph {et~al.},\ }\href@noop {} {\bibfield  {journal} {\bibinfo  {journal} {Nat. Phys.}\ }\textbf {\bibinfo {volume} {14}},\ \bibinfo {pages} {716} (\bibinfo {year} {2018})}\BibitemShut {NoStop}%
\bibitem [{\citenamefont {Zhang}\ \emph {et~al.}(2024)\citenamefont {Zhang}, \citenamefont {Hu}, \citenamefont {Ma}, \citenamefont {Li}, \citenamefont {Wang}, \citenamefont {Li}, \citenamefont {Movsesyan}, \citenamefont {Wang}, \citenamefont {Govorov}, \citenamefont {Gan} \emph {et~al.}}]{zhang2024quantum}%
  \BibitemOpen
  \bibfield  {author} {\bibinfo {author} {\bibfnamefont {C.}~\bibnamefont {Zhang}}, \bibinfo {author} {\bibfnamefont {H.}~\bibnamefont {Hu}}, \bibinfo {author} {\bibfnamefont {C.}~\bibnamefont {Ma}}, \bibinfo {author} {\bibfnamefont {Y.}~\bibnamefont {Li}}, \bibinfo {author} {\bibfnamefont {X.}~\bibnamefont {Wang}}, \bibinfo {author} {\bibfnamefont {D.}~\bibnamefont {Li}}, \bibinfo {author} {\bibfnamefont {A.}~\bibnamefont {Movsesyan}}, \bibinfo {author} {\bibfnamefont {Z.}~\bibnamefont {Wang}}, \bibinfo {author} {\bibfnamefont {A.}~\bibnamefont {Govorov}}, \bibinfo {author} {\bibfnamefont {Q.}~\bibnamefont {Gan}}, \emph {et~al.},\ }\href@noop {} {\bibfield  {journal} {\bibinfo  {journal} {Nat. Commun.}\ }\textbf {\bibinfo {volume} {15}},\ \bibinfo {pages} {2} (\bibinfo {year} {2024})}\BibitemShut {NoStop}%
\bibitem [{\citenamefont {Gatteschi}\ \emph {et~al.}(2006)\citenamefont {Gatteschi}, \citenamefont {Sessoli},\ and\ \citenamefont {Villain}}]{gatteschi2006molecular}%
  \BibitemOpen
  \bibfield  {author} {\bibinfo {author} {\bibfnamefont {D.}~\bibnamefont {Gatteschi}}, \bibinfo {author} {\bibfnamefont {R.}~\bibnamefont {Sessoli}},\ and\ \bibinfo {author} {\bibfnamefont {J.}~\bibnamefont {Villain}},\ }\href@noop {} {\emph {\bibinfo {title} {Molecular nanomagnets}}},\ Vol.~\bibinfo {volume} {5}\ (\bibinfo  {publisher} {Oxford University Press, USA},\ \bibinfo {year} {2006})\BibitemShut {NoStop}%
\bibitem [{\citenamefont {Jones}\ \emph {et~al.}(2000)\citenamefont {Jones}, \citenamefont {Vedral}, \citenamefont {Ekert},\ and\ \citenamefont {Castagnoli}}]{jones2000geometric}%
  \BibitemOpen
  \bibfield  {author} {\bibinfo {author} {\bibfnamefont {J.~A.}\ \bibnamefont {Jones}}, \bibinfo {author} {\bibfnamefont {V.}~\bibnamefont {Vedral}}, \bibinfo {author} {\bibfnamefont {A.}~\bibnamefont {Ekert}},\ and\ \bibinfo {author} {\bibfnamefont {G.}~\bibnamefont {Castagnoli}},\ }\href@noop {} {\bibfield  {journal} {\bibinfo  {journal} {Nature}\ }\textbf {\bibinfo {volume} {403}},\ \bibinfo {pages} {869} (\bibinfo {year} {2000})}\BibitemShut {NoStop}%
\bibitem [{\citenamefont {Zanardi}\ and\ \citenamefont {Rasetti}(1999)}]{zanardi1999holonomic}%
  \BibitemOpen
  \bibfield  {author} {\bibinfo {author} {\bibfnamefont {P.}~\bibnamefont {Zanardi}}\ and\ \bibinfo {author} {\bibfnamefont {M.}~\bibnamefont {Rasetti}},\ }\href@noop {} {\bibfield  {journal} {\bibinfo  {journal} {Physics Letters A}\ }\textbf {\bibinfo {volume} {264}},\ \bibinfo {pages} {94} (\bibinfo {year} {1999})}\BibitemShut {NoStop}%
\bibitem [{\citenamefont {Barron}(1986)}]{barron1986symmetry}%
  \BibitemOpen
  \bibfield  {author} {\bibinfo {author} {\bibfnamefont {L.}~\bibnamefont {Barron}},\ }\href@noop {} {\bibfield  {journal} {\bibinfo  {journal} {Chem. Soc. Rev.}\ }\textbf {\bibinfo {volume} {15}},\ \bibinfo {pages} {189} (\bibinfo {year} {1986})}\BibitemShut {NoStop}%
\bibitem [{\citenamefont {Tang}\ and\ \citenamefont {Cohen}(2010)}]{tang2010optical}%
  \BibitemOpen
  \bibfield  {author} {\bibinfo {author} {\bibfnamefont {Y.}~\bibnamefont {Tang}}\ and\ \bibinfo {author} {\bibfnamefont {A.~E.}\ \bibnamefont {Cohen}},\ }\href@noop {} {\bibfield  {journal} {\bibinfo  {journal} {Phys. Rev. Lett.}\ }\textbf {\bibinfo {volume} {104}},\ \bibinfo {pages} {163901} (\bibinfo {year} {2010})}\BibitemShut {NoStop}%
\end{thebibliography}%
\end{document}